\documentclass[aps,twocolumn,showkeys,nofootinbib,groupedaddress,superscriptaddress,showpacs]{revtex4-2}
\usepackage[utf8]{inputenc}
\usepackage{graphicx}
\usepackage{url}
\usepackage{amsmath}
\usepackage{color,soul}
\usepackage[parfill]{parskip}

\def\ie{\textit{i.e.}}
\def\eg{\textit{e.g.}}

\usepackage{amsmath}

\DeclareMathOperator*{\argmin}{arg\,min}

\begin{document}
\title{On the accuracy of short-term COVID-19 fatality forecasts}
\author{Nino Antulov-Fantulin}
\email{anino@ethz.ch}
\affiliation{Computational Social Science, ETH Zurich, 8092, Zurich, Switzerland}
\author{Lucas B\"ottcher}
\email{lucasb@g.ucla.edu}
\affiliation{Dept.~of Computational Medicine, University of California, Los Angeles, 90095-1766, Los Angeles, United States}
\affiliation{Computational Social Science, Frankfurt School of Finance and Management, Frankfurt am Main, 60322, Germany}
\date{\today}
\begin{abstract}
\begin{description}
\item[Background] Forecasting new cases, hospitalizations, and disease-induced deaths is an important part of infectious disease surveillance and helps guide health officials in implementing effective countermeasures. For disease surveillance in the U.S., the Centers for Disease Control and Prevention (CDC) combine more than 65 individual forecasts of these numbers in an ensemble forecast at national and state levels. 
\item[Methods] We collected data on CDC ensemble forecasts of COVID-19 fatalities in the United States, and compare them with easily interpretable ``Euler'' forecasts serving as a model-free benchmark that is only based on the local rate of change of the incidence curve. The term ``Euler method'' is motivated by the eponymous numerical integration scheme that calculates the value of a function at a future time step based on the current rate of change.
\item[Results] Our results show that CDC ensemble forecasts are not more accurate than ``Euler'' forecasts on short-term forecasting horizons of one week. However, CDC ensemble forecasts show a better performance on longer forecasting horizons. 
\item[Conclusions] Using the current rate of change in incidences as estimates of future incidence changes is useful for epidemic forecasting on short time horizons. An advantage of the proposed method over other forecasting approaches is that it can be implemented with a very limited amount of work and without relying on additional data (\eg, human mobility and contact patterns) and high-performance computing systems. 
\end{description}
\end{abstract}
%
\maketitle
Over the course of the COVID-19 pandemic more than 65 international research groups contributed to an ensemble forecast of reported COVID-19 cases, hospitalizations, and fatalities in the U.S.~\cite{CDCforecast}. These forecasts are a central source of information on the further development of the pandemic and used by various governmental and non-governmental entities including the Centers for Disease Control and Prevention (CDC)~\cite{Ray2020.08.19.20177493}. 

Different forecasting methods rely on different underlying models and assumptions. One may roughly divide forecasting models into three different classes: (i) mechanistic models~\cite{keeling2011modeling,bottcher2020unifying}, (ii) purely data-driven models~\cite{mills2019applied}, and (iii) hybrid models. Most classical epidemic models are mechanistic and aim at describing disease dynamics in terms of interacting individuals in a population. Such models are usually applied to describe the influence of certain factors (\eg, population density, demographics, contact patterns, mobility, etc.) on the dynamics of an epidemic. Data-driven or machine learning models make fewer assumptions about the underlying dynamics and are applicable to a broader range of forecasting problems, but they also come at the cost of less interpretability for policymakers and epidemiologists. 

Here, we show that a very basic, model-free forecasting approach provides effective short-term forecasts of COVID-19 fatalities. We refer to this method as ``Euler forecast'', owing to its mathematical connection to the Euler method~\cite{euler1794institutiones,quarteroni2010numerical} that is used in computational mathematics to calculate the value of a function at a future time step based on the current rate of change.
\begin{figure*}
    \centering
    \includegraphics{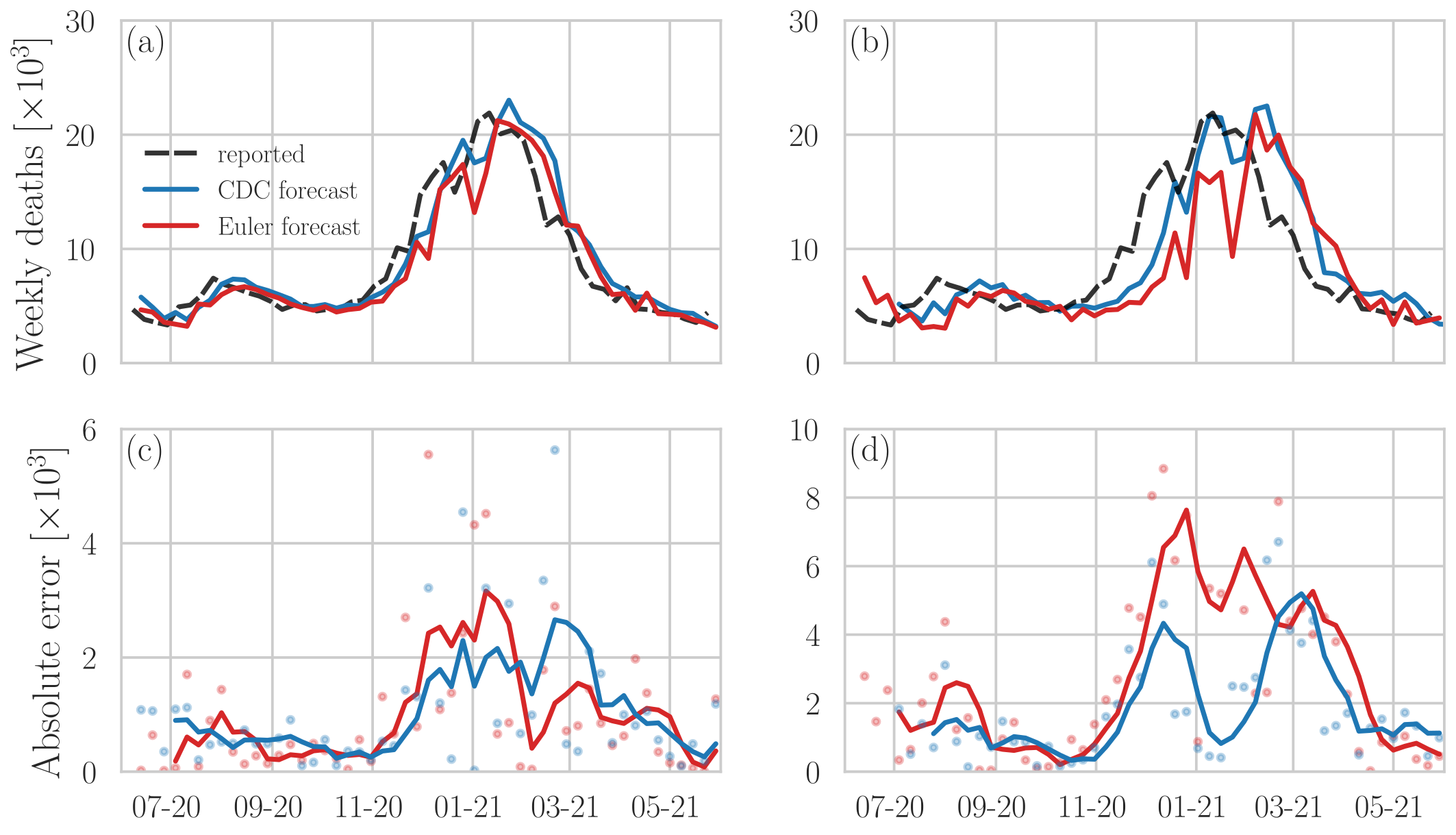}
    \caption{Comparison of predicted and reported weekly COVID-19 deaths in the U.S. (a,b) Forecasts of reported weekly COVID-19 deaths in the US for (a) 1-week and (b) 4-week forecasting horizons. Blue and red lines represent CDC ensemble forecasts~\cite{CDCforecast} and regularized Euler forecasts [Eq.~\eqref{eq:euler_regularized}] with $\lambda=10$, respectively. Reported COVID-19 fatalities (dashed black lines) are based on \cite{dong2020interactive}. (c,d) 4-week moving averages of weekly forecasting errors of Euler--Lagrange and CDC ensemble forecasts. Solid lines indicate 4-week moving averages that are calculated based on the shown data points.}
    \label{fig:comparison}
\end{figure*}
\section*{Methods}
We collected data on CDC ensemble forecasts between June 2020 and June 2021~\cite{CDCforecast}. Ensemble forecasts are available for cumulative and weekly incidence numbers and a forecasting horizon between one to four weeks. All forecasts use data from the Johns Hopkins Coronavirus Resource Center~\cite{dong2020interactive} as reference. Forecasts are made for epidemiological weeks which run Sunday through Saturday. As an example, if forecasts with one and four-week forecasting horizons are being made on June 7, 2020 the corresponding target dates are June 13, 2020 and July 4, 2020~\cite{CDCforecastData}.

We compare CDC ensemble forecasts of COVID-19 fatalities with a simple and easily interpretable forecasting method. To do so, let $y(t)$ be the incidence of COVID-19 fatalities at time $t$. We use $\dot{y}(t)$ to denote the rate of change of $y(t)$ at time $t$. Forecasting the incidence $y(t+\Delta t)$ at a target time $t+\Delta t$ requires us to find an estimate of this quantity at an earlier time $t$. A straightforward way to construct short-term forecasts is to use the current rate of change $\dot{y}(t)$ and determine a forecast at time $t_k=t_0+k\Delta t$ according to the Euler method~\cite{euler1794institutiones,quarteroni2010numerical}
\begin{equation}
y(t_0+k\Delta t)= \underbrace{y(t_0)}_{\text{last incidence}} + \underbrace{k\Delta t\,\dot{y}(t_0)}_{\text{incidence correction}},
\label{eq:Euler}
\end{equation}
where $\Delta t$ and $k=1,2,\dots$ represent a time step (\eg, one week) and the number of time steps in the forecasting horizon, respectively. 
However, observed incidences are subject to observation noise that results from confounding factors including sampling bias, measurement errors, and reporting delays~\cite{bottcher2021using}.

A possible way to ``de-noise'' observed data is to use previous weekly incidences instead of daily incidence levels. If observational noise can be reduced by averaging over a period of several days, daily errors are less pronounced on a weekly level. However, the local daily derivative is quite sensitive to noise and our incidence correction term is not helping in making accurate short-term forecasts. Therefore, we can impose some degree of regularity to reduce the level of noise with the following minimization
\begin{equation}
\argmin_{\{w_k\}} \sum_k (y_k-w_k)^2 + \lambda \sum_k (w_k-w_{k-1})^2\,,
\label{eq:total-variation}
\end{equation}
where $y_k=y(t_0+k\Delta t)$, $w_k=w(t_0+k \Delta t)$ is a regularized approximation of $y_k$, and $\lambda$ is a regularization parameter. In the limit $\lambda\rightarrow 0$, the argument of Eq.~\eqref{eq:total-variation} is minimized if $w(t)$ approaches $y(t)$. In the limit $\lambda\rightarrow\infty$, the argument of Eq.~\eqref{eq:total-variation} is minimized if $w(t)$ is constant (\ie, if $w_k-w_{k-1}=0$). This optimization process has its equivalent Euler--Lagrange formulation for differentiation~\cite{cullum1971numerical, chartrand2011numerical}. Values of $\lambda\in(0,\infty)$ yield functions $w(t)$ that are smoothened versions of $y(t)$ with respect to the discrete rate of change $w_k-w_{k-1}$. Finally, the regularized Euler short-term forecast\footnote{All optimization procedures in this work are applied in a causal manner. That is, at the prediction time $T$ only historical data $y(t\leq T)$ is being used in the minimization \eqref{eq:total-variation}.} is given by
\begin{equation}
y(t+k\Delta t)= y(t) + k\,[w(t)-w(t-\Delta t)]\,.   
\label{eq:euler_regularized}
\end{equation}
In the following section, we utilize the regularized Euler method to generate forecasts of reported COVID-19 fatalities. 

Our source codes are publicly available at \cite{GitRepoNinoCDC}.
\section*{Results}
Figure~\ref{fig:comparison} shows CDC ensemble forecasts (solid blue lines) of the weekly incidences of reported COVID-19 fatalities. The dashed black lines indicate reported COVID-19 fatalities. Between June and early November 2020, the majority of reported cases were close to the ensemble forecast. As COVID-19 deaths surged in November 2020, the forecasts of the ensemble method became less accurate than in previous months.

For a comparison between the CDC ensemble point estimates and those obtained with the regularized Euler method [Eq.~\eqref{eq:euler_regularized}], Fig.~\ref{fig:comparison} also shows Euler-method forecasts (solid red lines) of weekly incidences of COVID-19 fatalities in the U.S. 
We observe that one-week CDC ensemble forecast for the majority of data points are not more accurate than one-week Euler forecasts [Fig.~\ref{fig:comparison}(a)], which we use as a local-derivative-based forecasting benchmark. Although Euler and CDC forecasts still exhibit a similar structure for a four-week forecasting horizon [Fig.~\ref{fig:comparison}(b)], the Euler method is associated with larger deviations from the reported fatalities than the CDC ensemble method. To quantify differences in forecasting errors between the two methods, we use
\begin{equation}
\delta_{x,y}(t)=|x(t)-y(t)|
\label{eq:error1}
\end{equation}
to denote the absolute error between the Euler forecast $x(t)$ and CDC forecast $y(t)$ for target time $t$. 

Figure~\ref{fig:comparison}(c,d) show the 4-week moving averages of weekly forecasting errors $\delta(t)$ (solid lines) of the Euler (red) and CDC ensemble (blue) methods. As suggested by our above discussion of Fig.~\ref{fig:comparison}(a,b), we observe that the error of the Euler method is substantially smaller than that of the ensemble forecast for a one-week forecasting horizon. In about 61\% of the forecasting instances shown in Fig.~\ref{fig:comparison}(a), the regularized Euler method has a smaller error than the CDC ensemble forecast. The cumulative forecasting errors are 49,925 (Euler) and 52,885 (CDC). Without correction term [\ie, for $k=0$ in Eq.~\eqref{eq:euler_regularized}], the cumulative forecasting error of the Euler method is 52,660, again smaller than that of CDC ensemble forecast. Note that no regularization corresponds to a simple shift of the incidence curve. For a 4-week forecasting horizon [Fig.~\ref{fig:comparison}(d)], the cumulative error of the CDC ensemble forecast is 87,717, about 35\% smaller than that of the Euler method.
\section*{Discussion}
Our results suggest that easily interpretable methods like the Euler method, a model-free local-derivative-based forecasting benchmark, provide an effective alternative to more complex epidemic forecasting frameworks on short-term forecasting horizons. Similar conclusions were drawn in a recent study~\cite{katsikopoulos2021transparent} that compared Euler-like forecasts with those generated by Google Flu Trends. Regularized Euler forecasts have smaller errors with respect to CDC ensemble forecasts on one-week forecasting horizons in about 61\% of all cases. Simple curve shifts without regularization provide better one-week forecasts in 63\% of all cases, yet with a mean absolute error that is about 5\% larger than that found for regularized Euler forecasts. For longer forecasting horizons, it is not surprising that CDC forecasts that rely on additional input data, and epidemiological and statistical models become more accurate than Euler-like forecasting benchmarks.

One clear advantage of Euler forecasting methods is that they are less labor and resource intensive than more complex forecasting models, which often rely on the knowledge of expert groups and require specialized computing infrastructure. In their simplest implementation, Euler forecasts use the currently observed incidence rate as an estimate of the incidence rate in the following week. The regularization methods \eqref{eq:euler_regularized} can help further improve such data-driven forecasts.

In agreement with \cite{katsikopoulos2021transparent}, our results emphasize the importance of benchmarking complex forecasting models against simple forecasting baselines to further improve forecasting accuracy. Our study also points towards recent findings on algorithm rejection and aversion~\cite{dietvorst2020people} that found that ``\emph{people have diminishing sensitivity to forecasting error}'' and  that ``\emph{people are less likely to use the best possible algorithm in decision domains that are more unpredictable}''. Finally, in highly uncertain and noisy forecasting regimes, simple methods tend to outperform more complex methods because of a more favorable bias-variance tradeoff~\cite{friedman2001elements}. 
\bibliographystyle{unsrt}
\bibliography{refs.bib}
\end{document}